\documentclass[12pt]{article}
\usepackage{epsfig}
\usepackage{psfrag}
\usepackage{latexsym}
\usepackage{indentfirst}
\usepackage{fancyhdr}
\usepackage{amssymb}
\usepackage{amsfonts}
\usepackage{cite}
\usepackage{bbold}
\usepackage[footnotesize]{caption2}
\usepackage{graphicx}
\usepackage[center,footnotesize,hang]{subfigure}
\usepackage{url}
\usepackage{color}

\newcommand{\be}{\beta}

\newcommand{\beq}{\begin{equation}}
\newcommand{\eeq}{\end{equation}}
\newcommand{\bac}{\beq\begin{array}}
\newcommand{\eac}{\end{array}\eeq}
\newcommand{\ba}{\begin{array}}
\newcommand{\ea}{\end{array}}
\newcommand{\bea}{\begin{eqnarray}}
\newcommand{\eea}{\end{eqnarray}}
\newcommand{\beaa}{\begin{eqnarray*}}
\newcommand{\eeaa}{\end{eqnarray*}}

\def\beq{\begin{equation}}
\def\eeq{\end{equation}}
\def\bea{\begin{eqnarray}}
\def\eea{\end{eqnarray}}
\def\bet{\begin{tabular}}
\def\eet{\end{tabular}}
\def\bes{\begin{subequations}\bea}
\def\ees{\eea\end{subequations}}

\def\be{\begin{equation}}
\def\ee{\end{equation}}
\def\bc{\begin{center}}
\def\ec{\end{center}}
\def\bea{\begin{eqnarray}}
\def\eea{\end{eqnarray}}
\def\dd{\displaystyle}

\catcode`@=11
\def\marginnote#1{}
\newcount\hour
\newcount\minute
\newtoks\amorpm
\hour=\time\divide\hour by60
\minute=\time{\multiply\hour by60 \global\advance\minute by-\hour}
\edef\standardtime{{\ifnum\hour<12 \global\amorpm={am}%
        \else\global\amorpm={pm}\advance\hour by-12 \fi
        \ifnum\hour=0 \hour=12 \fi
        \number\hour:\ifnum\minute<10 0\fi\number\minute\the\amorpm}}
\edef\militarytime{\number\hour:\ifnum\minute<10 0\fi\number\minute}
\def\draftlabel#1{{\@bsphack\if@filesw {\let\thepage\relax
   \xdef\@gtempa{\write\@auxout{\string
      \newlabel{#1}{{\@currentlabel}{\thepage}}}}}\@gtempa
   \if@nobreak \ifvmode\nobreak\fi\fi\fi\@esphack}
        \gdef\@eqnlabel{#1}}
\def\@eqnlabel{}
\def\@vacuum{}
\def\draftmarginnote#1{\marginpar{\raggedright\scriptsize\tt#1}}
\def\draft{\oddsidemargin 0.0truein
        \def\@oddfoot{\sl preliminary draft \hfil
        \rm\thepage\hfil\sl\today\quad\militarytime}
        \let\@evenfoot\@oddfoot \overfullrule 3pt
        \let\label=\draftlabel
        \let\marginnote=\draftmarginnote
   \def\@eqnnum{(\theequation)\rlap{\kern\marginparsep\tt\@eqnlabel}%
\global\let\@eqnlabel\@vacuum}  }
\catcode`@=12
\begin{document}
\begin{titlepage}
\vspace*{-1cm}
\phantom{hep-ph/***}

\hfill{RM3-TH/10-13}
\hfill{CERN-PH-TH/2010-293}

\vskip 2.5cm
\begin{center}
{\Large\bf Different $SO(10)$ Paths to Fermion Masses and Mixings}

\end{center}
\vskip 0.2  cm
\vskip 0.5  cm
\begin{center}
{\large Guido Altarelli}~\footnote{e-mail address: guido.altarelli@cern.ch}
\\
\vskip .1cm
Dipartimento di Fisica `E.~Amaldi', Universit\`a di Roma Tre
\\
INFN, Sezione di Roma Tre, I-00146 Rome, Italy
\\
\vskip .1cm
and
\\
CERN, Department of Physics, Theory Division
\\
CH-1211 Geneva 23, Switzerland
\\

\vskip .2cm
{\large Gianluca Blankenburg}~\footnote{e-mail address: blankenburg@fis.uniroma3.it} 
\\
\vskip .1cm
Dipartimento di Fisica `E.~Amaldi', Universit\`a di Roma Tre
\\
INFN, Sezione di Roma Tre, I-00146 Rome, Italy
\\
\end{center}
\vskip 0.7cm
\begin{abstract}
\noindent
Recently $SO(10)$ models with type-II see-saw dominance have been proposed as a promising framework for obtaining Grand Unification theories with approximate Tri-bimaximal (TB) mixing in the neutrino sector. We make a general study of $SO(10)$ models with type-II see-saw dominance and show that an excellent fit can be obtained for fermion masses and mixings, also including the neutrino sector.  To make this statement more significant we compare the performance of type-II see-saw dominance models in fitting the fermion masses and mixings with more conventional models which have no built-in TB mixing in the neutrino sector. For a fair comparison the same input data and fitting procedure is adopted for all different theories. We find that the type-II dominance models lead to an excellent fit, comparable with the best among the available models, but the tight structure of this framework implies a significantly larger amount of fine tuning with respect to other approaches. 
\end{abstract}
\end{titlepage}
\setcounter{footnote}{0}
\vskip2truecm
%%%%%%%%%%%%%%%%%%%%%%%% 1.  INTRODUCTION   %%%%%%%%%%%%%%%%%%%%%%%%%%%%%%
%
\section{Introduction}

In this article we make a quantitative comparison of the performance of different types of $SO(10)$ Grand Unification Theories (GUT's) in reproducing the observed values of fermion masses and mixing, also including the neutrino sector.  By now we have a rather precise knowledge of the leptonic mixing angles \cite{review, rev2, Fogli, Maltoni} which, within the experimental accuracy, are consistent with the Tri-Bimaximal (TB) pattern \cite{hps}, \cite{RMP} and, as such, are very different from the quark mixing angles. A still open and challenging problem is that of formulating a natural model of Grand Unification based on $SO(10)$, leading not only to a good description of quark masses and mixing but also, in addition, of charged lepton masses and to approximate TB neutrino mixing. In $SO(10)$ the main added difficulty with respect to $SU(5)$ is clearly that all fermions in one generation belong to a single 16-dimensional representation, so that one cannot separately play with the properties of the $SU(5)$-singlet right-handed neutrinos in order to explain the striking difference between quark and neutrino mixing. There are a number of rather complete $SO(10)$ models, with different architectures,  that, without having a built-in TB mixing yet are able to reproduce the data on neutrino mixing angles
\cite{listSO(10), DR, AB1, AB2, JLM, JK, grim2}. These models fall in different classes: renormalizable or not, with lopsided or with symmetric mass matrices, with various assumed flavour symmetry, with different types of see-saw and so on.  In most of these models some dedicated parameters are available to fit the observed neutrino masses and mixing angles. In these models TB mixing appears as accidental, and if the data would become somewhat different, the new values of the mixing angles could as well be fitted by simply changing the values of the parameters. These models are certainly interesting and, in this article, a number of them will be confronted with the data and their respective performances will be compared and taken as a reference. A more difficult goal would be the construction of  $SO(10)$ models where TB mixing is built in and is automatic in a well defined first approximation, due, for example, to an underlying (broken) flavour symmetry. The leading approximation in these models is particularly constrained and the neutrino mixing angles are fixed in this limit. There are a number of GUT models of this type based on $SU(5)$ (see, for example, \cite{afh, Chen, ishi}), but, as mentioned, the $SO(10)$ case is more difficult and the existing attempts, in our opinion, are still not satisfactory in all respects. 

A promising strategy in order to separate charged fermions and neutrinos in $SO(10)$ is to assume the dominance of type-II see-saw \cite{tII} (with respect to type-I see-saw \cite{tI}) for the light neutrino mass matrix. Grand Unified $SO(10)$ models based on type-II see-saw dominance have been studied in refs. {\cite{babu,  brah, BSV, bajc1, goh, babu2, berto1, boh, mimura, DMM}. If type-II seesaw is responsible for neutrino
masses, then the neutrino mass matrix (proportional to) $f$ (see eqs.(\ref{eq1}, \ref{vr}, \ref{eq4} )) is separated from the dominant contributions to the charged fermion masses and can therefore show a completely different pattern. This is to be compared with the case of type-I see-saw where the neutrino mass matrix depends on the neutrino Dirac and Majorana matrices and, in $SO(10)$, the relation with the charged fermion mass matrices is tighter. Here we do not consider the problem of formulating a flavour symmetry or another dynamical principle that can lead to approximate TB mixing, but rather study the performance of the type-II see-saw $SO(10)$ models in fitting the data on fermion masses in comparison with other model architectures. Actually, we will show in Sect. 2 that, without loss of generality, we can always go to a basis where the matrix $f$ is of the TB type. In fact, since the TB mixing matrix is independent of the mass eigenvalues, the most general neutrino mass matrix, which is a symmetric complex matrix, can always be transformed into a TB mixing mass matrix by a change of the charged lepton basis. The observed deviations from TB mixing will then be generated by the diagonalisation of charged leptons and, in order to agree with the data, must be small. In turn the charged lepton mixings are related in $SO(10)$ to the quark mixings. Thus, in this class of models TB mixing is exact in the approximation of neglecting charged fermion mixings. When a symmetry guarantees TB mixing in first approximation the corrections from the diagonalisation of charged leptons are automatically small, while in general could be large. The main purpose of our analysis is to see to which extent this particular structure models is supported by the data among different types of $SO(10)$ .

In renormalizable $SO(10)$ models (a non necessary assumption which is only taken here in some cases for simplicity) the fermion
masses are generated by Yukawa couplings with Higgs fields transforming as {\bf 10}, ${\bf \overline{126}}$ (both symmetric) and {\bf 120} (antisymmetric).  Alternatively, in non renormalizable $SO(10)$ models the large representations  ${\bf\overline{126}}$  and {\bf 120} can be effectively obtained from the tensor products of smaller Higgs representations.
The {\bf 10} Yukawa couplings
contributing to up, down and charged lepton masses in most models have a large 33 term, corresponding to the large third generation masses, while all other entries are smaller and lead by themselves to zero CKM mixing (because the {\bf 10} contributes equally to up and down mixing). Quark mixings arise from small corrections due to
${\bf \overline{126}}$, the same Higgs representation that determines $f$ which, in models with type-II see-saw, is dominant in the neutrino sector, and to {\bf 120}.  Thus, in this approach, in the absence of  {\bf 120}, there is a strict relation between quark masses and mixings and the neutrino mass matrix. The presence of {\bf 120} dilutes this connection which however still remains important. In particular the small deviations from TB mixing induced by the diagonalisation of the charged lepton mass matrix, barring cancelations, are of the same order as the quark mixing angles and most of the parameters appear in both.  

An interesting question is to see to which extent the data are compatible with the constraints implied by this interconnected structure. A goal of this work is precisely to study how the inclusion of TB mixing along these lines is reflected in the ability of the model in reproducing the data in comparison with different structures. Thus, independent of the problem of determining a flavour symmetry that can fix the parameters to their required values,  we study $SO(10)$ models based on the dominance of type-II see-saw with respect to their performance in fitting the fermion masses and mixings in comparison with alternative realistic $SO(10)$ models. Some other analyses have appeared in the literature where $SO(10)$ models with dominance of type-II see-saw have been confronted with the data (see, for example, \cite{berto1, berto3, JK}). The present analysis is different and, in some respect, more general and, in addition, also includes the comparison with the most established models based on $SO(10)$ that can be considered both realistic (i.e. that have been worked out to a level that a comparison with the data is possible) and complete (i.e. that also include the neutrino sector). Each model will be compared with the same set of data on masses and mixings given at the GUT scale. For this purpose we specify a set of data for reference. We are not much concerned with the uncertainties which for sure exist in evolving the physical quantities measured at the electroweak scale up to the corresponding ones at the GUT scale. We will adopt a reasonable evolution up to the GUT scale, as can be found in the literature \cite{koide, parida, malinsky, berto3}, and consider the resulting set of data as the truth and fit all relevant models to that set (to be precise, we actually consider two sets of data, one for models with small $\tan{\beta}$ and one for those with large $\tan{\beta}$). We argue that if a particular model is  sizably better than another in fitting these idealized data it will also score better on the real data. For our comparison the first quality factor is the $\chi^2$ or the $\chi^2$/d.o.f. obtained from the fit for each model. We also introduce
 a parameter $d_{FT}$ for a quantitative measure of the amount of fine-tuning of parameters which is needed in each model. 
We find that $SO(10)$ models based on dominance of type-II see-saw can fit the data remarkably well, both in absolute terms and in comparison with other types of models, but at the price of a substantial amount of fine-tuning needed to compensate for the tension introduced by the double role of $f$ in determining both the neutrino mass matrix and the charged fermion masses and mixings. In particular the smallness of the first generation masses requires a precise cancellation of larger terms.

This article is organised as follows. In Sect. 2 we define our reference model, renormalizable with type-II dominance. In Sect. 3 we discuss the relations between parameters of the model and measured quantities. 
In Sect. 4 we specify the set of data that we adopt at the GUT scale. In Sect. 5 we describe our fitting procedure and the quality indicators that we consider to compare different models. In Sect. 6 we briefly describe a number of alternative $SO(10)$ models and we fit them to the same data as we did for our reference model. Finally in Sect. 7 we present our summary and conclusion. The best fit parameters and observables are listed in Appendix A for all the competing models.

\section{A class of $SO(10)$ models}

We consider the class of renormalizable Supersymmetric (SUSY) $SO(10)$ models with dominance of type-II see-saw for neutrino masses \cite{babu,  brah, BSV, bajc1, goh, babu2, berto1, boh, mimura, DMM}. In the following we indicate the generic model of this class by T-IID (from Type-II Dominance). In renormalizable $SO(10)$ models the Higgs fields that
contribute to fermion masses are in {\bf 10} (denoted by $H$), ${\bf \overline{126}}$
($\overline{\Delta}$) and {\bf 120} ($\Sigma$). 
The Yukawa
superpotential of this model is then given by:
\begin{eqnarray}
W_Y~=~h\, \psi\psi H + f\, \psi\psi\bar{\Delta}+h'\,\psi\psi \Sigma,
\label{eq1}
\end{eqnarray}
where the symbol $\psi$ stands for the {\bf 16} dimensional
representation of SO(10) that includes all the fermion fields in one generation.
The coupling matrices $h$ and $f$ are symmetric,
while $h^\prime$ is anti-symmetric.
The representations $H$ and $\Delta$ have two standard model
(SM) doublets in each of them whereas $\Sigma$ has four such doublets.
At the GUT scale $M_{GUT}$,
once the GUT and the $B-L$ symmetry are broken, one linear
combination of the up-type and one of down-type doublets
remain almost massless whereas the remaining combinations acquire GUT
scale masses.
The electroweak symmetry is broken after the light Minimal Supersymmetric Standard Model (MSSM) doublets
(to be called $H_{u,d}$) acquire vacuum expectation values (vevs) and
they then generate the
fermion masses. The resulting mass formulae for  different
fermion masses are given by (see, for example, \cite{mimura}):
\begin{eqnarray}
Y_u &=& h + r_2 f +r_3 h^\prime, \label{eq2} \\\nonumber
Y_d &=& r_1 (h+ f + h^\prime)\,, \\\nonumber
Y_e &=& r_1 (h-3f + c_e h^\prime)\,, \\\nonumber
Y_{\nu^D} &=& h-3 r_2 f + c_\nu h^\prime,
\end{eqnarray}
where $Y_a$ are mass matrices divided by the electro-weak vev's
$v_{u,d}$ and $r_a$ ($a=1,2,3$ and $c_b$ ($b=e,\nu$) are the mixing parameters which
relate the $H_{u,d}$ to the doublets in the various GUT
multiplets.

In generic $SO(10)$ models of this type, the neutrino
mass formula has a type-II \cite{tII} and a type-I \cite{tI}
contributions:
\begin{eqnarray}
{\cal M}_\nu~=~fv_L-M_D\frac{1}{fv_R}M^T_D\,,
\label{vr}
\end{eqnarray}
where $v_L$ is the vev of the $B-L=2$ triplet in the ${\bf\overline{126}}$ Higgs
field. Note that in general, the two
contributions to neutrino mass depend on two different parameters, $v_L$ and $v_R$,
and it is possible to have a symmetry breaking pattern in
$SO(10)$ such that the first contribution (the type-II term)
dominates over the type-I term. The possible realisation of this dominance and its consistency with coupling unification has been studied in the literature \cite{goh, aulak1, aulak2, bajc1} and found tricky but not impossible \cite{melfo}. The neutrino mass formula then
becomes
\begin{eqnarray}
{\cal M}_\nu~\sim~fv_L.
\label{eq4}
\end{eqnarray}
Note that $f$ is the same coupling matrix that appears in the
charged fermion masses in Eq. (\ref{eq2}),
up to factors from the Higgs mixings and the Clebsch-Gordan coefficients. Also note that the neutrino Dirac mass, proportional to $Y_{\nu^D}$ in eqs. \ref{eq2}, only enters in the neglected type-I see-saw terms and does not play a role in the following analysis.
The equations (\ref{eq2}) and (\ref{eq4}) are
the key equations in this approach.

For generic eigenvalues $m_i$, the most general matrix that is diagonalised by the TB unitary transformation is given by:
\beq
f=U_{TB}^* {\rm diag}(m_1,m_2,m_3) U_{TB}^\dagger~~~.
\label{numass}
\eeq
where, with a specific phase convention, we can take:
\begin{equation}
U_{TB}= \left(\matrix{
\dd\sqrt{\frac{2}{3}}&\dd\frac{1}{\sqrt 3}&0\cr
-\dd\frac{1}{\sqrt 6}&\dd\frac{1}{\sqrt 3}&-\dd\frac{1}{\sqrt 2}\cr
-\dd\frac{1}{\sqrt 6}&\dd\frac{1}{\sqrt 3}&\dd\frac{1}{\sqrt 2}}\right)~~~, 
\label{2}
\end{equation}
In this convention $U_{TB}$ is a real orthogonal matrix and all phases can be included in the eigenvalues $m_i$.  Then the matrix $f$ is symmetric with complex entries and,
from eq. \ref{numass}, one obtains:
\begin{equation}
f=\left(\matrix{
f_2&f_1&f_1\cr
f_1&f_2+f_0&f_1-f_0\cr
f_1&f_1-f_0&f_2+f_0}\right)~~~,
\label{effe}
\end{equation}
with: $m_1=f_2-f_1$, $m_2=f_2+2f_1$ and $m_3=f_2-f_1+2f_0$.

An important observation is that, for a generic neutrino mass matrix $f'$, we can always go to a basis where $f'$ is diagonalised by the TB unitary transformation in eq. \ref{2} and is of the form in eq. \ref{effe}. In fact, if we start from a complex symmetric matrix $f'$ not of that form, it is sufficient to diagonalise it by a unitary transformation $U$: $f'_{diag}=U^Tf'U$ and then take the matrix 
\begin{equation}
f=U_{TB}^*f'_{diag} U_{TB}^\dagger=U_{TB}^* U^T f' U U_{TB}^\dagger
\label{efefpr}
\end{equation} 
As a result the matrices $f$ and $f'$ are related by a change of the charged lepton basis induced by the unitary matrix $O= U U_{TB}^\dagger$ (in $SO(10)$ the matrix $O$ rotates the  whole fermion representations $\bf 16_i$).  Since TB mixing is a good approximation to the data we argue that this basis is a good starting point. In other words, for the physical neutrino mass matrix, TB mixing is a good approximation so that the unitary transformation $O$ is close to the identity. In this basis the deviations from TB mixing will be generated by the diagonalisation of charged leptons which, in order to agree with the data, must be small.  At the same time also the quark mixings must be small in order to correspond to the data.

In the selected basis, the parameterisations adopted for the matrices $h$ (symmetric) and $h^\prime$ (antisymmetric) are given by:
\begin{equation}
h=\left(\matrix{
h_{11}&h_{12}&h_{13}\cr
h_{12}&h_{22}&h_{23}\cr
h_{13}&h_{23}&Y}\right)~~~,
\label{h10}
\end{equation}
and
\begin{equation}
h^\prime=i\left(\matrix{
0&\sigma_{12}&\sigma_{13}\cr
-\sigma_{12}&0&\sigma_{23}\cr
-\sigma_{13}&-\sigma_{23}&0}\right)~~~,
\label{h120}
\end{equation}
The $h_{33}$ element of $h$ has been emphasised with the special notation $Y$ because usually it is the dominant term: $Y>>h_{ij}$. All the matrix elements $f_{ij}$, $Y$, $h_{ij}$ and $\sigma_{ij}$, as well as $r_a$ and $c_b$ that appear in eq. \ref{eq2}, will be taken as real. This leads to a crucial economy of parameters (justified by the fact that, as we shall see, the resulting fit is very good) and can be seen to correspond to an underlying "parity" symmetry \cite{boh} that implies that all mass matrices obtained from $h$, $h^\prime$ and $f$ are hermitian. Note that, due to the imaginary unit factor $i$ in front of the $h^\prime$ matrix, this purely imaginary matrix will in general induce CP violation in the quark and lepton sectors.

The charged fermion mass matrices are given by:
\begin{equation}
M_u=v_u\left(\matrix{
r_2f_2+h_{11}&r_2f_1+h_{12}+ir_3\sigma_{12}&r_2f_1+h_{13}+ir_3\sigma_{13}\cr
r_2f_1+h_{12}-ir_3\sigma_{12}&r_2(f_0+f_2)+h_{22}&r_2(-f_0+f_1)+h_{23}+ir_3\sigma_{23}\cr
r_2f_1+h_{13}-ir_3\sigma_{13}&r_2(-f_0+f_1)+h_{23}-ir_3\sigma_{23}&Y+r_2(f_0+f_2)}\right)~~~,
\label{mu}
\end{equation}
\begin{equation}
M_d=\frac{r_1v_u}{\tan\beta}\left(\matrix{
f_2+h_{11}&f_1+h_{12}+i\sigma_{12}&f_1+h_{13}+i\sigma_{13}\cr
f_1+h_{12}-i\sigma_{12}&f_0+f_2+h_{22}&-f_0+f_1+h_{23}+i\sigma_{23}\cr
f_1+h_{13}-i\sigma_{13}&-f_0+f_1+h_{23}-i\sigma_{23}&Y+f_0+f_2}\right)~~~,
\label{md}
\end{equation}
\begin{equation}
M_e=\frac{r_1v_u}{\tan\beta}\left(\matrix{
-3f_2+h_{11}&-3f_1+h_{12}+ic_e\sigma_{12}&-3f_1+h_{13}+ic_e\sigma_{13}\cr
-3f_1+h_{12}-ic_e\sigma_{12}&-3(f_0+f_2)+h_{22}&-3(-f_0+f_1)+h_{23}+ic_e\sigma_{23}\cr
-3f_1+h_{13}-ic_e\sigma_{13}&-3(-f_0+f_1)+h_{23}-ic_e\sigma_{23}&Y-3(f_0+f_2)}\right)~~~,
\label{me}
\end{equation}
where $\tan\beta=v_u/v_d$, with $v_{u,d}$ being the vacuum expectation values of $H_{u,d}$, and $M_u$, $M_d$ and $M_e$ refer to up, down and charged leptons, respectively. Note that all mass matrices are hermitian, hence each of them can be diagonalised by a unitary matrix. We have 6 real parameters in $h$, 3 in $h'$, 3 in $f$ plus $r_1/tan\beta$, $r_2$, $r_3$, $c_e$ and the ratio $v_L/v_u$, or a total of 17 parameters. The (in principle) measurable quantities are 12 fermion masses, 6 mixing angles, 2 CP violating phases (we do not include Majorana phases) or a total of 20. Only 18 of these observables have been measured (the PNMS phase and one of the neutrino masses are unconstrained). 
In the following more predictive versions of the model with less free parameters will also be considered (for example, the JK2 model in Sect. 6). 

\section{Observables and Parameters}

Starting from the neutrino mass matrix in eqs. \ref{eq4}, \ref{effe} we obtain:
\begin{equation}
m_{\nu1}=(f_2-f_1)v_L,~~~m_{\nu2}=(f_2+2f_1)v_L,~~~m_{\nu3}=(f_2-f_1+2f_0)v_L.
\label{mnui}
\end{equation}
The oscillation frequencies are given by:
\begin{eqnarray}
\Delta m^2_{sol}&=&m_{\nu2}^2-m_{\nu1}^2=3f_1(f_1+2f_2)v_L^2\label{freq} \\\nonumber
\Delta m^2_{atm}&=&m_{\nu3}^2-m_{\nu1}^2=4f_0(f_0+f_2-f_1)v_L^2
\end{eqnarray}
and their ratio $r$ is:
\begin{equation}
r=\frac{\Delta m^2_{sol}}{\Delta m^2_{atm}}=\frac{3f_1(f_1+2f_2)}{4f_0(f_0+f_2-f_1)}
\label{defr}
\end{equation}
The experimental smallness of $r \sim 1/30$ suggests that $f_1/f_0$ is small, $|f_1/f_0| \sim 0.1-0.2$  (barring an accidental cancellation with $f_1 \sim -2f_2$).

The deviations of the leptonic mixing angles from the TB values are induced by the diagonalisation of the charged lepton mass matrix given in eq. \ref{me}. As $M_e$ is hermitian it is diagonalised by a unitary transformation:
\be
M_e = U_e M_e^{diag} U_e^{\dagger} 
\label{diagme}
\ee
where $M_e^{diag}$ stands for a diagonal matrix with real non-negative elements $m_e$, $m_\mu$, $m_\tau$ and $U_e$ is the relevant unitary matrix. Any unitary matrix can in general be written as (see, for example, \cite{afm}):
\be
U = e^{i \phi_0} {\rm diag}(e^{i(\phi_1 + \phi_2)}, e^{i \phi_2}, 1)\tilde U 
                {\rm diag}(e^{i(\phi_3 + \phi_4)}, e^{i \phi_4}, 1) ~,
                 \label{uni}
\ee
where $\phi_i$ (i=0,...,4) run from $0$ to $2 \pi$ and $\tilde U$ is the standard parameterization for the CKM
mixing matrix, namely
\be
\tilde U~=~ 
\left(\matrix{1&0&0 \cr 0&c_{23}&s_{23}\cr0&-s_{23}&c_{23}     } 
\right)
\left(\matrix{c_{13}&0&s_{13}e^{i\delta} \cr 0&1&0\cr -s_{13}e^{-i\delta}&0&c_{13}     } 
\right)
\left(\matrix{c_{12}&s_{12}&0 \cr -s_{12}&c_{12}&0\cr 0&0&1     } 
\right)~~~,
\label{uffi}
\ee 
where all the mixing angles belong to the first quadrant and $\delta$ to $[0,2 \pi]$. An approximate form of $m_e$ that follows from eq. \ref{diagme} for $m_\tau>>m_\mu>>m_e$, in a linear approximation in the small mixing angles unless non linear terms are rescued by large mass factors, is given by:
\begin{equation}
M_e=\left(\matrix{
m_e+m_\mu s_{12}^{e2}+m_\tau s_{13}^{e2}&(m_\mu s_{12}^e+m_\tau s_{13}^e s_{23}^e e^{i \delta^e})e^{i \phi^e_1}&m_\tau s_{13}^e e^{i (\delta^e+\phi^e_1+\phi^e_2)}\cr
(m_\mu s_{12}^e+m_\tau s_{13}^e s_{23}^e e^{-i \delta^e})e^{-i \phi^e_1}&m_\mu+m_\tau s_{23}^{e2}&m_\tau s_{23}^e e^{i \phi^e_2}\cr
m_\tau s_{13}^e e^{-i (\delta^e+\phi^e_1+\phi_2^e)}&m_\tau s_{23}^e e^{-i \phi^e_2}&m_\tau }\right)~~~.
\label{meprime}
\end{equation}
Going back to eq.\ref{me} and  similar ones for the heavier generations, in the above approximation, we can write down simple analytic expressions for the mass eigenvalues (except for the first generation mass which is fine tuned in this class of models) and the mixing angles and phases. Note, however, that exact expressions are used in our numerical fits. Starting from the 2-3 family sector we have:
\begin{eqnarray}
m_\tau&\approx& k[Y-3(f_0+f_2)]~~~\label{obs23} \\\nonumber
m_\mu&\approx& k [h_{22}-3(f_0+f_2)]-m_\tau s_{23}^{e2}\\\nonumber
s_{23}^e e^{i \phi^e_2}&\approx& \frac{k}{m_\tau}[h_{23}+3(f_0-f_1)+ ic_e \sigma_{23 }].
\label{eigenv}
\end{eqnarray}
with
\begin{equation}
k= \frac{r_1 v_u}{\tan\beta}
\label{kappa}
\end{equation}
We also obtain:
\begin{equation}
s_{13}^e e^{i (\delta^e+\phi^e_1+\phi^e_2)} \approx  \frac{k}{m_\tau}[h_{13}-3 f_1+ ic_e \sigma_{13 }]
\label{s13}
\end{equation}
The following expression for $s_{12}^e$ is less accurate but, in general, still sufficiently good, at least for indicative purposes. It is obtained from the 11 entries of the matrices in eqs. \ref{me}, \ref{meprime} by neglecting $m_e$:
\begin{equation}
s_{12}^e  \approx \sqrt{\frac{k}{m_\mu}(h_{11}-3f_2) - \frac{m_\tau}{m_\mu}s_{13}^{e2}}
\label{s12}
\end{equation}
A remaining phase can be derived from the equation (obtained from the 12 entries of the matrices in eqs. \ref{me}, \ref{meprime}):
\begin{equation}
(m_\mu s_{12}^e+m_\tau s_{13}^e s_{23}^e e^{i \delta^e})e^{i \phi^e_1} \approx k(h_{12}-3f_1+ ic_e \sigma_{12 })
\label{phie1}
\end{equation}
From fitting the data one finds that indeed $Y$ is the largest parameter, followed by $h_{13}$, $h_{22}$, $h_{23}$, $\sigma_{13}$, $\sigma_{23}$ and $f_0$ while $f_1$, $f_2$, $h_{11}$, $h_{12}$ and $\sigma_{12}$ are still smaller. 

In a linear approximation in the leptonic mixing angles $s_{ij}^e$ the corrections to $U_{TB}$ from the diagonalization of charged leptons are given by (for each matrix element we omit an overall phase):
\begin{eqnarray}
U_{12} &\approx& \frac{1}{\sqrt{3}} (1-s_{12}^e e^{i\phi^e_1}-s_{13}^e e^{i(\delta^e + \phi^e_1 +\phi^e_2)})~~~\label{uij} \\\nonumber
U_{13} &\approx& \frac{1}{\sqrt{2}}  (s_{12}^e - s_{13}^e e^{i(\delta^e + \phi^e_2)}) \\\nonumber
U_{23} &\approx& \frac{-1}{\sqrt{2}}  (1 + s_{23}^e e^{i\phi^e_2}).
\end{eqnarray}
The corresponding corrected mixing angles are:
\begin{eqnarray}
\sin{\theta_{12}}=|U_{12}| &\approx &\frac{1}{\sqrt{3}} (1- s_{12}^e \cos{\phi^e_1} -  s_{13}^e \cos{(\delta^e + \phi^e_1 +\phi^e_2)}) ~~~\label{tetaij} \\\nonumber
\theta_{13}=|U_{13}| &\approx & \frac{1}{\sqrt{2}} \sqrt{s_{12}^{e2}+s_{13}^{e2}-2s_{12}^e s_{13}^e \cos{(\delta^e + \phi^e_2)}}  \\\nonumber
\sin{\theta_{23}}=|U_{23}| &\approx &\frac{1}{\sqrt{2}} (1+  s_{23}^e \cos{\phi^e_2}). 
\end{eqnarray}
It turns out, however, that the above linear approximations are often not sufficiently accurate because the leptonic mixing angles $s_{ij}^{e}$ are not small enough. This is particularly true for $\sin{\theta_{12}}$ and $\sin{\theta_{13}}$. So the above linearised formulae are only given for indicative purposes and, in our fit, we used the exact expressions for the mixing, obtained from $U^\dagger_e U_{TB}$ from eqs. \ref{uni}, \ref{uffi}. But the approximate analytic formulae are useful to understand the need of fine tuning to reproduce the observed masses for the light generations of leptons and the neutrino masses and mixing. We see in fact from eqs. (\ref{freq}, \ref{defr}) that neutrino masses impose a strong constraint on the values of $f_i$.  But the same $f_i$ enter in the charged lepton masses and the leptonic mixing angles eqs. (\ref{eigenv}, \ref{s13}, \ref{s12}) and they must conspire with the $h_{ij}$ and $\sigma_{ij}$ parameters in order to reproduce the observed values. And the same is also true for the quark masses that we now discuss.

In the quark sector the approximate formulae for the 2-3 families, analoguous to eqs. \ref{eigenv} for charged leptons, are derived from eqs. \ref{mu}, \ref{md}:
\begin{eqnarray}
m_b&\approx& k[Y+f_0+f_2]~~~\label{obsd23} \\\nonumber
m_s&\approx& k [h_{22}+f_0+f_2]-m_b s_{23}^{d2}\\\nonumber
s_{23}^d e^{i \phi^d_2}&\approx& \frac{k}{m_\tau}[h_{23}-f_0+f_1+ i \sigma_{23 }].
\label{eigenvd}
\end{eqnarray}
\begin{eqnarray}
m_t&\approx& v_u[Y+r_2(f_0+f_2)] ~~~\label{obsu23}\\\nonumber
m_c&\approx& v_u [r_2f_0+r_2f_2+h_{22}]-m_t s_{23}^{u2}\\\nonumber
s_{23}^u e^{i \phi^u_2}&\approx& \frac{v_u}{m_\tau}[h_{23}+r_2(f_1-f_0)+ i r_3\sigma_{23 }].
\end{eqnarray}
Note that, as $Y>>|f_i|$,  $m_b=m_\tau$ approximately holds at the GUT scale (bottom-tau unification).
Relations analogous to eqs. \ref{s13}, \ref{s12}, \ref{phie1} can be readily written down for the up and down quark sectors. For the CKM mixing matrix we go back to eq. \ref{uni} and derive $U_u$ and $U_d$ that diagonalise $m_u$ and $m_d$, given in eqs. \ref{mu}, \ref{md}, respectively. We then construct $V_{CKM}=U_u^\dagger U_d$.

\section{A set of idealized data at the GUT scale}

For the charged fermion masses and CKM mixings at the GUT scale we used the values given in Tab.\ref{tab:ferm-carichi}\subref{tab:tgbeta10}-Tab.\ref{tab:ferm-carichi}\subref{tab:tgbeta55} as input for our fitting procedure. These values are based on the analysis of ref.\cite{parida}, in the MSSM framework, and were obtained from a two-loop Renormalization Group Evolution (RGE) from the SUSY scale at about 1 $TeV$ to the GUT scale $\sim 2 \times 10^{16}$ $GeV$. Ref.\cite{parida} can be considered as an upgrade of a previous work \cite{koide}. This analysis has been repeated for two different values of the supersymmetric (SUSY) parameter $\tan{\beta}$: $\tan{\beta} = 10$ and $\tan{\beta} = 55$. In fact $\tan{\beta}$ is the most important SUSY parameter that directly affects the RGE. So these tables have been used to fit two typical classes of models with small or large $\tan{\beta}$. Actually for both values of $\tan{\beta}$ we have used in the fit an updated version of the data which was derived in refs.\cite{berto3}, \cite{malinsky} shown in Tab.\ref{tab:ferm-carichi}. In particular, refs.\cite{berto3}, \cite{malinsky} use more recent values for $m_u$, $m_d$ and $m_s$ at low energy with respect to the ones used in ref.\cite{parida}. The errors taken on the data points at the GUT scale are those shown in ref.\cite{berto3} or in ref.\cite{malinsky} and do not include an estimate of the additional ambiguities associated with the chosen procedure. For example, the SUSY spectrum parameters, which mainly enter in the evolution through the threshold corrections, are subject to considerable ambiguities \cite{threshold}. Moreover at the GUT scale other threshold corrections appear, but they are again affected by large uncertainties and model dependent. As we are only interested in defining a set of idealized data in order to make a meaningful performance test for a number of different models,  we decided to ignore the uncertainties on both the SUSY- and GUT-scale threshold corrections.  As already discussed we are not much concerned with these uncertainties, rather we prefer to fit all models on the same set of data, arguing that if a particular model is better than another in fitting these idealized data it would also score better on the real data if they were known.

\begin{table}[htbp]
\begin{center}
\subtable[$tg\beta=10$ \label{tab:tgbeta10}]{
\begin{tabular}{|l|r|}
\hline
Observables & Input data \\
\hline
\hline
$m_u [MeV]$ & $0.55 \pm 0.25$\\
\hline
$m_c [MeV]$ & $210 \pm 21$\\
\hline
$m_t$ [GeV]& $82.4^{+30.3}_{-14.8}$\\
\hline
$m_d [MeV]$ & $1.24 \pm 0.41$\\
\hline
$m_s [MeV]$ & $21.7 \pm 5.2$\\
\hline 
$m_b [GeV]$ & $1.06^{+0.14}_{-0.09}$\\
\hline
$m_e [MeV]$ & $0.3585 \pm 0.0003$\\
\hline
$m_\mu [MeV] $ & $75.672 \pm 0.058$\\
\hline
$m_\tau [GeV]$ & $1.2922 \pm 0.0013$\\
\hline
$V_{us}$ & $0.2243 \pm 0.0016$\\
\hline
$V_{cb}$ & $0.0351 \pm 0.0013$\\
\hline
$V_{ub}$ & $0.0032 \pm 0.0005$\\
\hline
$J\times 10^{-5}$ & $2.2 \pm 0.6$\\
\hline
\end{tabular}
%\end{center}
%\caption{GUT scale data for charged fermions for $tg\beta=10$}
%\label{tab:tgbeta10} %
}\qquad\qquad
%\end{table}
%%%%%%%%%%%%%%%%%%%%%%%%%%%%%%%
%\begin{table}[htbp]
%\begin{center}
\subtable[$\tan\beta = 55$ \label{tab:tgbeta55}]{
\begin{tabular}{|l|r|}
\hline
Observables & Values \\
\hline
\hline
$m_u [MeV]$ & $0.45 \pm 0.2$\\
\hline
$m_c [MeV]$ & $217 \pm 35$\\
\hline
$m_t [GeV]$ & $97 \pm 38$\\
\hline
$m_d [MeV]$ & $1.3 \pm 0.6$\\
\hline
$m_s [MeV]$ & $23 \pm 6$\\
\hline
$m_b [GeV]$ & $1.4 \pm 0.6$\\
\hline
$m_e [MeV]$ & $0.3565 \pm 0.001$\\
\hline
$m_\mu [MeV]$ & $75.3 \pm 0.12$\\
\hline
$m_\tau [GeV]$ & $1.629 \pm 0.037$\\
\hline
$V_{us}$ & $0.2243 \pm 0.0016$\\
\hline
$V_{cb}$ & $0.0351 \pm 0.0013$\\
\hline
$V_{ub}$ & $0.0032 \pm 0.0005$\\
\hline
$J\times 10^{-5}$ & $2.2 \pm 0.6$\\
\hline
\end{tabular}
}
\end{center}
\caption{GUT scale data for charged fermions for $tg\beta=10$ (ref.\cite{berto3}) and $\tan\beta=55$ (ref.\cite{malinsky})}
\label{tab:ferm-carichi}
\end{table}

Concerning the neutrino sector, we have ignored the effects of the evolution from the low energy scale to the GUT scale. In fact for all the models analysed here the mass spectrum is non degenerate, so that the evolution can be considered negligible to a good approximation \cite{neutrino-ev}. The low energy data are taken from \cite{Maltoni} and the corresponding values are given in Tab. \ref{tab:neutrino}.

\begin{table}[htbp]
\begin{center}
\begin{tabular}{|l|r|}
\hline
Observable & Input data \\
\hline
\hline
$\Delta m^2_{21}\times 10^{-5} [eV^2]$ & $7.65 \pm 0.23$ \\
\hline
$\Delta m^2_{31}\times 10^{-3} [eV^2]$ & $2.40 \pm 0.12$ \\
\hline
$sin^2\theta_{13}$& $0.010 \pm 0.016$ \\
\hline
$sin^2\theta_{12}$& $0.304 \pm 0.022$ \\
\hline
$sin^2\theta_{23}$& $0.50 \pm 0.07$ \\
\hline
\end{tabular}
\end{center}
\caption{Neutrino masses and mixing ( \cite{Maltoni})}
\label{tab:neutrino}
\end{table}

\section{Fitting procedure and quality factors}

As already discussed, the class of models T-IID described in Sect. 2 contains a total of 17 independent parameters. We are going to compare these models with the set of 18 "measurements" described in the previous section. As shown in eq. \ref{md}, \ref{me} the parameter $\tan{\beta}$ always enters in the combination $r_1/\tan{\beta}$ in all the observables. So in these models it is possible to obtain a large ratio $m_t/m_b$ without making $\tan{\beta}$  large (in fact, large $\tan{\beta}$ can be problematic in the MSSM with a GUT Yukawa $t-b-\tau$ unification \cite{Yuk-unification}). Here we are going to fit the model presented in Sect. 2 on the data in Tab.\ref{tab:ferm-carichi}\subref{tab:tgbeta10}, which refer to the $\tan{\beta}=10$ case. We use the numerical minimization tool Minuit2 developed at CERN.

We introduce a parameter $d_{FT}$ for a quantitative measure of the amount of fine-tuning needed in the models. This adimensional quantity is obtained as the sum of the absolute values of the ratios between each parameter and its "error", defined for this purpose as the shift from the best fit value that changes $\chi^2$ by one unit, with all other parameters fixed at their best fit values (this is not the error given by the fitting procedure because in that case all the parameters are varied at the same time and the correlations are taken into account):
\begin{equation}
\label{fine-tuning}
d_{FT} = \sum \mid \frac{par_i}{err_i} \mid
\end{equation}
It is clear that $d_{FT}$ gives a rough idea of the amount of fine-tuning involved in the fit because if some $|err_i/par_i|$ are very small it means that it takes a minimal variation of the corresponding parameters to make a large difference on $\chi^2$. The value of $d_{FT}$ for our best fit output is shown in Tab.\ref{tab:fit-model1}\subref{tab:fit-model1-obs}.
To get a better idea of the significance of this number, one can compare it with a similar number $d_{Data}$ based on the data, i.e. the sum of the absolute values of the ratios between each observable and its error as derived from the input data:
\begin{equation}
\label{data-precision}
d_{Data} = \sum \mid \frac{obs_i}{err_i} \mid
\end{equation}
In particular for the set of data in Tab.\ref{tab:ferm-carichi}\subref{tab:tgbeta10} $d_{Data}\sim 3800$ and for Tab.\ref{tab:ferm-carichi}\subref{tab:tgbeta55} $d_{Data}\sim 1300$.

The best fit results for the models T-IID are shown in Tab.\ref{tab:fit-model1}. In Tab.\ref{tab:fit-model1}\subref{tab:fit-model1-obs} we indicate the values of the observables obtained at the GUT scale, the $\chi^2$ calculated in each sector (quarks, charged leptons, neutrinos), the  $\chi^2$/d.o.f. (in the case of T-IID we have one degree of freedom, so the reduced $\chi^2$/d.o.f. is the same as the $\chi^2$) and the fine-tuning parameter $d_{FT}$. In Tab.\ref{tab:fit-model1}\subref{tab:fit-model1-par} we show the best fit parameters.
 
\begin{table}[htbp]
\begin{center}
\subtable[\label{tab:fit-model1-obs}]{%
\begin{tabular}{|l|r|}
\hline
Observable & Best fit value \\
\hline
\hline
$m_u [MeV]$ & 0.553 \\
\hline
$m_c [MeV]$ & 210 \\
\hline
$m_t [GeV]$ & 82.6 \\
\hline
$m_d [MeV]$ & 1.15 \\
\hline
$m_s [MeV]$ & 22.4 \\
\hline 
$m_b [GeV]$ & 1.08 \\
\hline
$m_e [MeV]$ & 0.3585 \\
\hline
$m_\mu [MeV] $ & 75.67 \\
\hline
$m_\tau [GeV]$ & 1.292 \\
\hline
$V_{us}$ & 0.224 \\
\hline
$V_{cb}$ & 0.0351 \\
\hline
$V_{ub}$ & 0.00320 \\
\hline
$J\times 10^{-5}$ & 2.19 \\
\hline
$\Delta m^2_{21} \times 10^{-5} [eV^2]$ & 7.65 \\
\hline
$\Delta m^2_{32} \times 10^{-3} [eV^2]$ & 2.40 \\
\hline
$sin^2\theta_{13} $ & 0.0126 \\
\hline
$sin^2\theta_{12} $ & 0.305 \\
\hline
$sin^2\theta_{23} $ & 0.499 \\
\hline\hline
$\chi^2$ quark & 0.0959 \\
\hline
$\chi^2$ charged fermions & 0.0959 \\
\hline
$\chi^2$ neutrino & 0.0316 \\
\hline
$\chi^2$ totale & 0.127 \\
\hline
$\chi^2/dof$ totale & 0.127 \\
\hline
$d_{FT}$ & 469777 \\
\hline
\end{tabular}
}\qquad\qquad
\subtable[\label{tab:fit-model1-par}]{
\begin{tabular}{|l|r|}
\hline
Parameter & Best fit value \\
\hline
\hline
$h_{11}v_u[GeV]$ & 0.808 \\
\hline
$h_{12}v_u[GeV]$ & 1.17 \\
\hline
$h_{13}v_u[GeV]$ & 6.06 \\
\hline
$h_{22}v_u[GeV]$ & 5.37 \\
\hline
$h_{23}v_u[GeV]$ & 5.64 \\
\hline
$Yv_u[GeV]$ & 85.0 \\
\hline
$f_0v_u[GeV]$ & -2.20 \\
\hline
$f_1v_u[GeV]$ & -0.276 \\
\hline
$f_2v_u[GeV]$ & -0.228 \\
\hline
$\sigma_{12}v_u[GeV]$ & -0.270 \\
\hline
$\sigma_{13}v_u[GeV]$ & 2.27 \\
\hline
$\sigma_{23}v_u [GeV]$ & 6.37 \\
\hline
$r_1/\tan{\beta}$ & 0.0129 \\
\hline
$r_2$ & 1.66 \\
\hline
$r_3$ & 0.612 \\
\hline
$c_e$ & 3.85 \\
\hline
$v_L/v_u \times 10^{-9}$ & 0.0112 \\
\hline
\end{tabular}
}
\end{center}
\caption{Fit result for the model T-IID described in Sect. 2}
\label{tab:fit-model1}
\end{table}

We caution that, due to the non linearity of the problem and the large number of parameters, many local minima are present in all the fits we have performed. Although we have carefully tested the selected minimum for possible improvements, still we cannot be sure that it indeed is the global minimum, within reasonable ranges for the parameters. But, since we obtain an excellent agreement with the data, we are not much concerned with this problem. In fact the resulting $\chi^2\sim 0.13$ is very good. We note however that a substantial level of fine tuning is needed. We in fact obtain $d_{FT} \sim 4.7~10^5$ from the fit, to be compared with $d_{Data} \sim 3.8~10^3$. As explained in Sect. 3, this is due to the fact that the neutrino oscillation frequency data impose strong constraints on the $f_i$ values. Those also enter in the expressions of the mass values and the mixings of charged fermions and in the deviations from TB neutrino mixing. As a result, the strong suppression of the first generation masses and the observed values of mixing angles can only be obtained by a fine tuning among the $f_i$ parameters with those of the {\bf10} and {\bf120} matrices.

An important conclusion that we can already give at this stage is that models of the T-IID type formulated in Sect. 2 in the general framework of type-II see-saw dominance can lead at a very good fit of the data but at the price of a pronounced level of fine tuning. In the following section we will compare this class of models with some other "realistic" $SO(10)$ models in the literature. 

\section{Comparison with other $SO(10)$ models}

In order to appreciate the performance of the above class of models in fitting the fermion masses and mixings we present a comparison with some other realistic $SO(10)$ 
models present in literature. For a meaningful comparison we applied the same fitting procedure to each model, using the same set of data as described in the previous section. A list of realistic $SO(10)$ theories can be found in \cite{listSO(10)}. Here we consider a number of models, with different structure and type (renormalizable, non renormalizable, lopsided or symmetric, with type-I and/or type-II see-saw and so on).

The model introduced and discussed in a series of papers by Dermisek and Raby (DR) \cite{DR} is an example of non-renormalizable $SO(10)$ theory, with Higgs multiplets in the {\bf 10}, {\bf 45} and ${\bf \overline{16}}$, based on the flavour symmetry $S_3\times U(1) \times Z_2 \times Z_2$. This model can be considered as a descendant of the model by Barbieri et al, \cite{Barbieri}, with the two lightest generations in a doublet of $U(2)$ and the third generation in a singlet. In fact, $S_3$, the permutation group of 3 objects, is a discrete subgroup of $SO(3)$ (and $SU(2)$ is the covering group of $SO(3)$) with inequivalent irreducible representations {\bf 2}, {\bf 1} and  {\bf 1'}. In the $S_3$ symmetry limit only the third generation masses are allowed. Then the second generation masses are generated by a symmetry breaking stage with intensity proportional to a parameter $\epsilon$, and finally the first generation masses only arise when an additional stage with intensity $\epsilon'$ is switched on. The magnitudes of $\epsilon$ and $\epsilon'$ are determined by a Froggatt-Nielsen mechanism, which is induced by a set of heavy fields that are then integrated away. In the neutrino sector a type-I see-saw mechanism with a hierarchical Majorana mass matrix is adopted. New $SO(10)$-singlet neutrino and scalar fields are introduced, so that, enough freedom is allowed that is essential to reproduce the observed neutrino properties. In this model the ratio of top to bottom masses is of order $\tan{\beta}$, which must be large: $m_t(m_t)/m_b(m_t) \sim \tan{\beta} \sim$  50.

A good realisation of the lopsided idea is given by the model by Albright, Babu and Barr  (ABB) \cite{AB1} \cite{AB2}, also non-renormalizable and with type-I see-saw. In minimal $SU(5)$ the down quark and the charged lepton mass matrices are connected by a transposition, as the roles of ${\bf \bar{5}}$ and {\bf10} are interchanged in the respective mass matrices. So, if the mass matrices are asymmetric (lopsided), left-handed mixings of charged leptons can be large without implying large left-handed mixings for quarks (in fact, this only implies large right-handed quark mixings which are not observable). In turn large charged lepton left-handed mixings contribute to the observed large neutrino mixings. In the ABB model the breaking of $SO(10)$ in fact preserves $SU(5)$ in a first stage, which makes this connection with lopsidedness relevant. This model is based on a flavour symmetry $U(1)\times Z_2 \times Z_2$, which, however, plays a different role than in the DR model. In fact, while in the DR model the flavour symmetry and its breaking are mainly used to reproduce the hierarchy of fermion masses and mixings, in the ABB construction, the main goal is to select the Lagrangian terms which are desired and reject those that would not reproduce the data. In particular, in the Higgs sector the symmetry is crucial in order to implement and preserve the mechanism of Dimopoulos and Wilczek for the solution of the doublet-triplet splitting problem \cite{BarrRaby}. As is often the case in non renormalizable models, the Higgs sector is inspired by a minimality requirement that demands the smallest possible representations compatible with realistic properties. Accordingly the Higgs sector contains {\bf 10}, ${\bf 16 + \overline{16}}$, {\bf 45} representations and a few $SO(10)$ singlets. The ABB model was originally formulated when the neutrino frequencies and mixing angles were not as precisely measured as now. Later the lepton sector has been revised in ref. \cite{AB2} with some ad hoc arbitrary ingredients and we have adopted this last description here as it can be fitted to the present data.

We have also considered a variation of this model proposed by Ji, Li, Mohapatra (JLM) \cite{JLM} which was motivated by a less ad-hoc treatment of the neutrino sector. Precisely, the down quark and the charged lepton mass matrices are the same as in ABB, while the up and Dirac neutrino mass matrices are modified by introducing some new vertices. The structure of neutrino mixing is not attributed to the Majorana matrix, which in this model is diagonal, but rather to the modified Dirac matrix. The model is again based on type-I see-saw and the new added operators introduce a sufficient number of new free parameters to accommodate the neutrino mixing angles.

Turning now to renormalizable models we have analysed the model referred to as BSV in Tab.\ref{tab:compared}. In this version one is introducing the minimal Higgs content in the Yukawa sector ({\bf 10}, ${\bf\overline{126}}$). This minimal model has been discussed in ref. \cite{BSV} and compared with the data in ref. \cite{berto3}  (where also the cases of type-I and mixed type-I and type-II were considered).  A different perspective on this model  is presented in ref. \cite{dor}, also including some comparison with the data (with mixed type-I and type-II see-saw). With this restricted Higgs content one cannot impose the L-R parity that leads to real hermitian $h$ and $f$ matrices, otherwise there is no CP violation and the too restricted number of parameters does not allow a good fit. Thus in ref. \cite{berto3} complex $h$ and $f$ matrices were taken. As a consequence there are more parameters than observables. We have repeated the fit of that model within our procedure and the results are listed in Tab.\ref{tab:compared}. We see from our fit in Tab.\ref{tab:compared} that, in spite of this multitude of parameters, with type-II dominance, in the absence of  the {\bf 120}, no good fit of the data can be obtained. 

Another particular class of renormalizable models with type-II see-saw dominance has been discussed by Joshipura and Kodrani (JK) \cite{JK}. This model comes in two versions, one with type-I dominance (JK1) and one with type-II dominance (JK2).  The Higgs coupled to fermions are in {\bf 10}, ${\bf\overline{126}}$ and {\bf 120}, like in the T-IID model. The characteristic feature of the model is the presence of a broken $\mu-\tau$ symmetry (an explicit breaking is present in the {\bf 10} mass terms) in addition to the parity symmetry which leads to hermitian mass matrices. In the case of type-II see-saw dominance this model is a particular case of T-IID with some restrictions on the parameters imposed by the ansatz of broken $\mu-\tau$ symmetry. 

The renormalizable model proposed by Grimus and Kuhbock \cite{grim2} is of particular interest for us because it closely corresponds to the T-IID model (the fermions masses arise from Higgs in {\bf 10}, ${\bf\overline{126}}$ and {\bf 120} and the parity symmetry of ref. \cite{boh} is assumed) except for the fact that it is based on type-I see-saw dominance. There is one more parameter, $v_R$ (see eq.(\ref{vr})), than in the model T-IID. The only other difference is that this model was fitted in the basis where the matrix $h$ from the {\bf 10} is diagonal and real, while for the T-IID we worked in the TB basis for the matrix $f$ of the ${\bf\overline{126}}$.

The model by Dermisek and Raby is the only one that demands a large value of $\tan{\beta}$, so we fit it on the data in Tab.\ref{tab:ferm-carichi}\subref{tab:tgbeta55}, while all other models are fitted on the values in Tab.\ref{tab:ferm-carichi}\subref{tab:tgbeta10}. A collection of our results on comparing the different models is shown in Tab.\ref{tab:compared}. In Appendix A we show the mass matrices of the different models in terms of the corresponding parameters and the results of the fitting procedure.

\begin{table}[htbp]
\begin{center}
\begin{tabular}{|l|l|l|l|l|l|}
\hline
Model &d.o.f. &$\chi^2$ &  $\chi^2$/d.o.f. & $d_{FT}$ & $d_{Data}$ \\
\hline
\hline
DR    \cite{DR} & 4 &0.41 & 0.10 &7.0 ~$10^3$&1.3~$10^3$\\
\hline
\hline
ABB    \cite{AB1, AB2, BarrRaby} &6 & 2.8 & 0.47 &8.1~$10^3$& 3.8~$10^3$\\
\hline
JLM    \cite{JLM} &4 &2.9 & 0.74 & 9.4~$10^3$& 3.8~$10^3$ \\
\hline
\hline
BSV    \cite{berto3} &$< 0$ &6.9 &-& 2.0~$10^5$& 3.8~$10^3$  \\
\hline
JK2   \cite{JK} &3 &3.4 & 1.1 &4.7~$10^5$& 3.8~$10^3$  \\
\hline
GK    \cite{grim2} &0 &0.15 & -& 1.5~$10^5$& 3.8~$10^3$  \\
\hline
T-IID    &1 & 0.13 & 0.13 & 4.7~$10^5$& 3.8~$10^3$ \\
\hline
\end{tabular}
\end{center}
\caption{Comparison of different $SO(10)$ models fitted to the data. The double lines mark three sectors: the DR model, non renormalizable with type-I see-saw and the only one with large $\tan{\beta}$,  the non renormalizable models ABB and JLM, lopsided with type-I see-saw and the renormalizable models BSV, JK2, T-IID, with type-II see-saw, and GK, renormalizable with type-I see-saw}
\label{tab:compared}
\end{table}

The results in Tab.\ref{tab:compared} are in some cases somewhat different than those from the fits described by the authors of the various models. One obvious reason is that the set of the input data for the fit is different, as explained in Sect. 4.
In some cases (e.g. for JK2 and GK) the difference also comes from the fact that, for hermitian matrices, we fitted the eigenvalues of $m_i$, that is of the mass matrices and not those of $m_i^\dagger m_i$. Indeed, from fitting the squares of masses one can end up with solutions of negative mass which we discard.

From the results in Tab.\ref{tab:compared} we see that the most established realistic $SO(10)$ models, DR, ABB, JLM, which are non renormalizable with type-I see-saw, achieve a $\chi^2$/d.o.f. smaller than 1 with a moderate level of fine tuning, defined by the parameter $d_{FT}$. In the above list of models DR is special as it has large $\tan{\beta}$, so that it was fitted to a different set of data, given in the right panel of Tab.\ref{tab:ferm-carichi}, which were obtained by evolving up to the GUT scale with large $\tan{\beta}$. The model T-IID which we have introduced and described in Sect. 2, realizes an excellent fit, but with a level of fine tuning considerably larger than in the DR, ABB and JLM models. As already mentioned, this large fine tuning arises from the difficulty of fitting the light 1st generation charged fermion masses, together with the neutrino oscillation frequencies and mixing angles. In fact the neutrino oscillation frequencies and mixing angles lead to $f$ matrix elements of comparable size that need cancelations to occur with the parameters in  $h$ and $h'$ in order to reproduce the light quark and lepton masses. Note that in the DR, ABB, JLM the fine tuning is less pronounced because in all these models new parameters appear in the neutrino sector, so that the neutrino masses and mixings are more independent from the charged fermion sectors. In the JK2 model the constraints from the  broken $\mu-\tau$ symmetry reduce the number of parameters within the general framework of the T-IID model. As a consequence the quality of the fit worsens and the level of fine tuning is the same. Instead the poor result of the BSV model, both in terms of $\chi^2$ and of $d_{FT}
$ shows that the presence of the {\bf 120} is crucial \cite{berto1, goh}. In ref. \cite{aulak2, bajc1, berto3, fuk} it was also shown that the contribution from the {\bf 120} is essential for consistency with the existing proton decay bounds. Finally the GK model shows that if one takes the same framework of T-IID, except that the assumption of type-I see-saw dominance is made, an excellent fit is also obtained, still with a substantial level of fine tuning.

We conclude this section by recalling that in this work we have compared the models only on the basis of their ability to fit the fermion masses and mixings. Clearly for a model to be complete and satisfactory many more aspects are important, like the mechanism of the GUT symmetry breaking, the running of the couplings towards the GUT unification, the compatibility of the model with the proton decay bounds and so on (see ref. \cite{aulak2} and refs therein). Also, for the models based on type-II see-saw dominance we assumed here that the dominance is absolute, while in reality one should estimate the corrections to this approximation or include the normal type-I see-saw terms \cite{berto3}. Additional quality factors could be considered. In models with an underlying flavour symmetry the smallness of some of the parameters is guaranteed by the broken flavour symmetry. Thus an important quality factor is the percentage of small quantities that are predicted to be small. For some of the models there is a broken flavour symmetry that reproduces, at least in part, the observed hierarchies. But this is not the case for T-IID where  the problem of an underlying flavour symmetry was not addressd here.

\section{Summary and conclusion}

We have made a quantitative comparison of the performance of different types of $SO(10)$ Grand Unification Theories (GUT's) in reproducing the observed values of fermion masses and mixing, also including the neutrino sector. All models, chosen among the most complete and sufficiently realistic, have been confronted with the same set of data (except for DR that requires a large value of $\tan{\beta}$), using the same fitting procedure. We have shown that a $SO(10)$ model with type-II see-saw dominance can achieve a very good fit of fermion masses and mixings also including the neutrino sector (provided that the representations {\bf 10}, ${\bf\overline{126}}$ and {\bf 120} are all included).  The quality of the fit in terms of $\chi^2$ and $\chi^2$/d.o.f. is comparable with the best realistic $SO(10)$ model that we have tested. However, the tight structure of the T-IID model implies a significantly larger amount of fine tuning with respect to more conventional models like the DR or the ABB and JLM models. But those models have no built-in TB mixing and in fact could accommodate a wide range of mixing angle values. A model with type-II see-saw dominance can offer a convenient framework for obtaining a GUT model with approximate TB mixing in the neutrino sector. For this goal a suitable flavour symmetry should be introduced in order to enforce TB mixing as a first approximation. Such a flavour symmetry construction was attempted in ref.\cite{DMM} but the resulting model is very sketchy and can only be considered as a first step. In fact, when the model of ref.\cite{DMM} is submitted to our fitting procedure it leads to $\chi^2\sim 342$, $\chi^2$/d.o.f $\sim 38$ and $d_{FT} \sim 1.2~10^4$ which is quite far from the performance of all models that we considered. The formulation of a natural and elegant model along these lines is not an easy task and more work is demanded. 

\section*{Acknowledgements}
We recognize that this work has been partly supported by the Italian Ministero dell'Universit\`a e della Ricerca Scientifica, under the COFIN program (PRIN 2008).
We thank Luca Merlo for many interesting comments and discussions.

\vfill
\newpage

\section*{Appendix A}

In this section we show the Yukawa matrices for every model that we have analysed together with the parameter values obtained from our fitting procedure. $M_{u/d/e/\nu}$ are the Dirac matrices for up-quarks/down-quarks/charged leptons/neutrinos. $M_{R/L}$ are the Majorana matrices for the right/left neutrinos.  

\begin{itemize}

\item Model DR

\begin{minipage}[c]{.5\textwidth}
\centering
\begin{eqnarray}
%\label{DR-Yuk-U}
M_u &=& \left(
\begin{array}{ccc}
0 & \epsilon^\prime\,\rho & -\epsilon\,\varepsilon \\
-\epsilon^\prime\,\rho & \tilde\epsilon\,\rho & -\epsilon \\
\epsilon\,\varepsilon & \epsilon & 1 \\
\end{array}
\right) \: \lambda  \sin{\beta} \nonumber\\
%\label{DR-Yuk-D}
M_d &=& \left(
\begin{array}{ccc}
0 & \epsilon^\prime & -\epsilon\,\varepsilon\,\sigma \\
-\epsilon^\prime & \tilde\epsilon & -\epsilon\,\sigma \\
\epsilon\,\varepsilon & \epsilon & 1 \\
\end{array}
\right) \: \lambda  \cos{\beta} \nonumber\\ 
%\label{DR-Yuk-L}
M_e &=& \left(
\begin{array}{ccc}
0 & -\epsilon^\prime & 3\,\epsilon\,\varepsilon \\
-\epsilon^\prime & 3\,\tilde\epsilon & 3\,\epsilon \\
-3\,\epsilon\,\varepsilon\,\sigma & -3\,\epsilon\,\sigma & 1 \\
\end{array}
\right) \: \lambda  \cos{\beta} \nonumber\\
%\label{DR-Yuk-N}
M_\nu &=& \left(
\begin{array}{ccc}
0 & -\epsilon^\prime\,\omega & \frac{3}{2}\,\epsilon\,\varepsilon\,\omega \\
-\epsilon^\prime\,\omega & 3\,\tilde\epsilon\,\omega & \frac{3}{2}\,\epsilon\,\omega \\
-3\,\epsilon\,\varepsilon\,\sigma & -3\,\epsilon\,\sigma & 1 \\
\end{array}
\right) \: \lambda  \sin{\beta} \nonumber\\
%\label{DR-Maj-R}
M_R &=& \left(
\begin{array}{ccc}
M_{R_1} & 0 & 0 \\
0 & M_{R_2} & 0 \\
0 & 0 & M_{R_3} \nonumber\\
\end{array}
\right) 
\end{eqnarray}
\end{minipage}
\begin{minipage}[c]{.5\textwidth}
\centering
\begin{tabular}{|l|r|}
\hline
Parameter & Best fit value\\
\hline\hline
$\lambda$ & 0.464\\
\hline
$\lambda \epsilon$ & -0.0215\\
\hline
$|\sigma|$ & 0.256\\
\hline
$\delta_\sigma$ & 0.120\\
\hline
$|\rho|$ & 0.0565\\
\hline
$\delta_\rho$ & -1.59\\
\hline
$\lambda|\tilde{\epsilon}|$ & 0.00656\\
\hline
$\delta_{\tilde{\epsilon}}$ & -0.57\\
\hline
$\lambda\epsilon^\prime$ & -0.00157\\
\hline
$\lambda|\varepsilon|$ & 0.00286\\
\hline
$\delta_\varepsilon$ & -2.45\\
\hline
$M_{R_1} \times 10^{13}[GeV]$ & $0.000226$\\
\hline
$M_{R_2} \times 10^{13}[GeV]$ & $0.0144$\\
\hline
$M_{R_3} \times 10^{13}[GeV]$ & $-17.5$\\
\hline
\end{tabular}
\end{minipage}

As specified by the authors of the model  we have taken $\tan{\beta}=50.34$. Some of the parameters are complex and, in these cases, we have set $ z=|z|e^{i\delta_{z}}$ and all the phases are expressed in radiants.
The quantity $\omega$ is given by $\omega = \frac{2\sigma}{2\sigma-1}$.

\newpage 
\item Model ABB

\begin{minipage}[c]{.5\textwidth}
\centering
\begin{eqnarray}
M_u&=&\left(
\begin{array}{ccc}
\eta & 0 & 0 \\
0 & 0 & \epsilon / 3 \\
0 & -\epsilon / 3 & 1
\end{array}
\right) M_U \nonumber\\
M_d&=&\left(
\begin{array}{ccc}
0 & \delta & \delta^\prime e^{i\phi} \\
\delta & 0 & \sigma + \epsilon / 3 \\
\delta^\prime e^{i\phi} & -\epsilon / 3 & 1
\end{array}
\right) M_D \nonumber\\
M_e&=&\left(
\begin{array}{ccc}
0 & \delta & \delta^\prime e^{i\phi} \\
\delta & 0 & -\epsilon \\
\delta^\prime e^{i\phi} & \sigma + \epsilon & 1
\end{array}
\right) M_D \nonumber\\
M_\nu&=&\left(
\begin{array}{ccc}
\eta & 0 & 0 \\
0 & 0 & -\epsilon \\
0 & \epsilon& 1
\end{array}
\right) M_U \nonumber\\
M_R &=& \left(
\begin{array}{ccc}
c^2\eta^2 & -b\epsilon\eta & a\eta \\
-b\epsilon\eta & \epsilon^2 & -\epsilon \\
a\eta & -\epsilon & 1
\end{array}
\right)\Lambda_R \nonumber
\end{eqnarray}
\end{minipage}
\begin{minipage}[c]{.5\textwidth}
\centering
\begin{tabular}{|l|r|}
\hline
Parameter & Best fit value\\
\hline\hline
$\epsilon$ & 0.140\\
\hline
$M_U [GeV]$ & 95.5\\
\hline
$\eta$ & 0.00000576\\
\hline
$\sigma$ & 1.72\\
\hline
$M_D [GeV]$ & 0.612\\
\hline
$\delta$ & 0.00795\\
\hline
$\delta^\prime$ & -0.00836\\
\hline
$\phi$ & -1.12\\
\hline
$a$ & -1.48\\
\hline
$b$ & -2.55\\
\hline
$c$ & 2.95\\
\hline
$\Lambda_{R} \times 10^{14}[GeV]$& 2.04\\
\hline
\end{tabular}
\end{minipage}
%in this case the $\sin\beta(\cos\beta)$ are included in the factors $M_{U(D)}$
\item Model JLM

\begin{minipage}[c]{.5\textwidth}
\centering
\begin{eqnarray}
M_u&=&\left(
\begin{array}{ccc}
\eta & 0 & k+\rho/3 \\
0 & 0 & \omega \\
k-\rho/3 & \omega & 1
\end{array}
\right) M_U \nonumber\\
M_d&=&\left(
\begin{array}{ccc}
0 & \delta & \delta^\prime e^{i\phi} \\
\delta & 0 & \sigma + \epsilon / 3 \\
\delta^\prime e^{i\phi} & -\epsilon / 3 & 1
\end{array}
\right) M_D \nonumber\\
M_e&=&\left(
\begin{array}{ccc}
0 & \delta & \delta^\prime e^{i\phi} \\
\delta & 0 & -\epsilon \\
\delta^\prime e^{i\phi} & \sigma + \epsilon & 1
\end{array}
\right) M_D \nonumber\\
M_\nu&=&\left(
\begin{array}{ccc}
\eta & 0 & k-\rho \\
0 & 0 & \omega \\
k+\rho & \omega & 1
\end{array}
\right) M_U \nonumber\\
M_R&=&\left(
\begin{array}{ccc}
a & 0 & 0 \\
0 & b & 0 \\
0 & 0 & 1
\end{array}
\right) \Lambda_R \nonumber
\end{eqnarray}
\end{minipage}
\begin{minipage}[c]{.5\textwidth}
\centering
\begin{tabular}{|l|r|}
\hline
Parameter & Best fit value\\
\hline\hline
$\epsilon$ & 0.144\\
\hline
$M_U [GeV]$ & 87.6\\
\hline
$\eta$ & $0.00000739$\\
\hline
$\sigma$ & 1.81\\
\hline
$M_D [GeV]$ & 0.589\\
\hline
$\delta$ & 0.00991\\
\hline
$\delta^\prime$ & 0.0141\\
\hline
$\phi$ & 0.468\\
\hline
$\omega$ & -0.0451\\
\hline
$\rho$ & 0.00899\\
\hline
$k$ & 0.0188\\
\hline
$a$ & -0.00159\\
\hline
$b$ & -0.00193\\
\hline
$\Lambda_{R} \times 10^{13}[GeV]$ & 2.07\\
\hline
\end{tabular}
\end{minipage}

\newpage

\item Model BSV

\begin{minipage}[c]{\textwidth}
\centering
\begin{eqnarray}
M_u&=&v_u\left(
\begin{array}{ccc}
r_2f_2+h_{11}&r_2f_1+h_{12}&r_2f_1+h_{13}\\
r_2f_1+h_{12}&r_2(f_0+f_2)+h_{22}&r_2(-f_0+f_1)+h_{23}\\
r_2f_1+h_{13}&r_2(-f_0+f_1)+h_{23}&Y+r_2(f_0+f_2)
\end{array}
\right) \nonumber\\
M_d&=&\frac{r_1v_u}{\tan\beta}\left(
\begin{array}{ccc}
f_2+h_{11}&f_1+h_{12}&f_1+h_{13}\\
f_1+h_{12}&f_0+f_2+h_{22}&-f_0+f_1+h_{23}\\
f_1+h_{13}&-f_0+f_1+h_{23}&Y+f_0+f_2
\end{array}
\right) \nonumber\\
M_e&=&\frac{r_1v_u}{\tan\beta}\left(
\begin{array}{ccc}
-3f_2+h_{11}&-3f_1+h_{12}&-3f_1+h_{13}\\
-3f_1+h_{12}&-3(f_0+f_2)+h_{22}&-3(-f_0+f_1)+h_{23}\\
-3f_1+h_{13}&-3(-f_0+f_1)+h_{23}&Y-3(f_0+f_2)\\
\end{array}
\right) \nonumber\\
 M_L &=& \left(
\begin{array}{ccc}
f_2&f_1&f_1\\
f_1&f_2+f_0&f_1-f_0\\
f_1&f_1-f_0&f_2+f_0
\end{array}
\right)v_L \nonumber
\end{eqnarray}
\end{minipage}

\begin{minipage}[c]{\textwidth}
%\centering
\begin{tabular}{|l|r|}
\hline
Parameter & Best fit value\\
\hline\hline
$|h_{11}|v_u[GeV]$ & 1.42 \\
\hline
$\delta_{h_{11}}$& -0.496\\
\hline
$|h_{12}|v_u[GeV]$ & 0.435 \\
\hline
$\delta_{h_{12}}$& 3.09\\
\hline
$|h_{13}|v_u[GeV]$ & 10.7 \\
\hline
$\delta_{h_{13}}$& -0.614\\
\hline
$|h_{22}|v_u[GeV]$ & 0.793 \\
\hline
$\delta_{h_{22}}$& 1.22\\
\hline
$|h_{23}|v_u[GeV]$ & 3.20 \\
\hline
$\delta_{h_{23}}$& 2.80\\
\hline
$|Y|v_u[GeV]$ & 79.9 \\
\hline
$\delta_{Y}$& -0.732\\
\hline
$f_0v_u[GeV]$ & -2.13 \\
\hline
$f_1v_u[GeV]$ & 0.369 \\
\hline
$f_2v_u[GeV]$ & 0.110 \\
\hline
$r_1/\tan{\beta}$ & 0.0148 \\
\hline
$|r_2|$ & 0.515 \\
\hline
$\delta_{r_{2}}$& 1.52\\
\hline
$v_L/v_u \times 10^{-9}$ & 0.0108 \\
\hline
\end{tabular}
\end{minipage}

Here again the complex parameters are understood as $z=|z|e^{i\delta_{z}}$ and the phases are in radiants.

\newpage

\item Model JK2

\begin{minipage}[c]{\textwidth}
\centering
\begin{eqnarray}
M_u&=&\left(
\begin{array}{ccc}
rh_{11}+sf_{11} & rh_{12}+sf_{12}+itg_{12} & rh_{12}+sf_{12}-itg_{12} \\
rh_{12}+sf_{12}-itg_{12} & rh_{22}+sf_{22} & rh_{23}+sf_{23}+itg_{23}  \\
rh_{12}+sf_{12}+itg_{12} & rh_{23}+sf_{23}-itg_{23} & rh_{33}+sf_{22}
\end{array}
\right)v \nonumber\\
M_d&=&\left(
\begin{array}{ccc}
h_{11}+f_{11} & h_{12}+f_{12}+ig_{12} & h_{12}+f_{12}-ig_{12} \\
h_{12}+f_{12}-ig_{12} & h_{22}+f_{22} & h_{23}+f_{23}+ig_{23}  \\
h_{12}+f_{12}+ig_{12} & h_{23}+f_{23}-ig_{23} & h_{33}+f_{22}
\end{array}
\right)v \nonumber\\
M_e&=&\left(
\begin{array}{ccc}
h_{11}-3f_{11} & h_{12}-3f_{12}+ipg_{12} & h_{12}-3f_{12}-ipg_{12} \\
h_{12}-3f_{12}-ipg_{12} & h_{22}-3f_{22} & h_{23}-3f_{23}+ipg_{23}  \\
h_{12}-3f_{12}+ipg_{12} & h_{23}-3f_{23}-ipg_{23} & h_{33}-3f_{22}
\end{array}
\right)v \nonumber\\
 M_L &=& \left(
\begin{array}{ccc}
f_{11} & f_{12} & f_{12} \\
f_{12} & f_{22} & f_{23}  \\
f_{12} & f_{23} & f_{22}
\end{array}
\right)r_L \nonumber
\end{eqnarray}
\end{minipage}

\begin{minipage}[c]{\textwidth}
%\centering
\begin{tabular}{|l|r|}
\hline
Parameter & Best fit value\\
\hline\hline
$h_{11}v[GeV]$ & 0.00204\\
\hline
$h_{22}v[GeV]$ & 0.576\\
\hline
$h_{23}v[GeV]$ & 0.120\\
\hline
$h_{33}v[GeV]$ & 0.619\\
\hline
$f_{11}v[GeV]$ & -0.000960\\
\hline
$f_{12}v[GeV]$ & -0.00398\\
\hline
$f_{22}v[GeV]$ & -0.0282\\
\hline
$f_{23}v[GeV]$ & 0.0381\\
\hline
$g_{12}v[GeV]$ & 0.00514\\
\hline
$g_{23}v[GeV]$ & 0.522\\
\hline
$r$ & 74.0\\
\hline
$s$ & 141\\
\hline
$t$ & 71.6\\
\hline
$p$ & 1.17\\
\hline
$r_L/v\times 10^{-9}$ & -0.741\\
\hline
\end{tabular}
\end{minipage}
\newline
\newline

\newpage

\item Model GK

\begin{minipage}[c]{\textwidth}
\centering
\begin{eqnarray}
M_u&=&\left(
\begin{array}{ccc}
r_Hh_{11}+r_Ff_{11} & r_Ff_{12}+ir_ug_{12} & r_Ff_{13}+ir_ug_{13} \\
r_Ff_{12}-ir_ug_{12} & r_Hh_{22}+r_Ff_{22} & r_Ff_{23}+ir_ug_{23}  \\
r_Ff_{13}-ir_ug_{13} & r_Ff_{23}-ir_ug_{23} & r_Hh_{33}+r_Ff_{33}
\end{array}
\right)v \nonumber\\
M_d&=&\left(
\begin{array}{ccc}
h_{11}+f_{11} & f_{12}+ig_{12} & f_{13}+ig_{13} \\
f_{12}-ig_{12} & h_{22}+f_{22} & f_{23}+ig_{23}  \\
f_{13}-ig_{13} & f_{23}-ig_{23} & h_{33}+f_{33}
\end{array}
\right)v \nonumber\\
M_e&=&\left(
\begin{array}{ccc}
h_{11}-3f_{11} & -3f_{12}+ir_lg_{12} & -3f_{13}+ir_lg_{13} \\
-3f_{12}-ir_lg_{12} & h_{22}-3f_{22} & -3f_{23}+ir_lg_{23}  \\
-3f_{13}-ir_lg_{13} & -3f_{23}-ir_lg_{23} & h_{33}-3f_{22}
\end{array}
\right)v \nonumber\\
 M_{\nu} &=& \left(
\begin{array}{ccc}
r_Hh_{11}-3r_Ff_{11} & -3r_Ff_{12}+ir_Dg_{12} & -3r_Ff_{13}+ir_Dg_{13} \\
-3r_Ff_{12}-ir_Dg_{12} & r_Hh_{22}-3r_Ff_{22} & -3r_Ff_{23}+ir_Dg_{23}  \\
-3r_Ff_{13}-ir_Dg_{13} & -3r_Ff_{23}-ir_Dg_{23} & r_Hh_{33}-3r_Ff_{33}
\end{array}
\right)v \nonumber\\
M_R&=&\left(
\begin{array}{ccc}
f_{11} & f_{12} & f_{13} \\
f_{12} & f_{22} & f_{23}  \\
f_{13} & f_{23} & h_{33}
\end{array}
\right)\frac{1}{r_R}
\end{eqnarray}
\end{minipage}

\begin{minipage}[c]{\textwidth}
%\centering
\begin{tabular}{|l|r|}
\hline
Parameter & Best fit value\\
\hline\hline
$h_{11}v[GeV]$ & 2.91\\
\hline
$h_{22}v[GeV]$ & 36.4\\
\hline
$h_{33}v[GeV]$ & 1130\\
\hline
$f_{11}v[GeV]$ & -1.40\\
\hline
$f_{12}v[GeV]$ & 0.779\\
\hline
$f_{13}v[GeV]$ & 6.29\\
\hline
$f_{22}v[GeV]$ & -15.1\\
\hline
$f_{23}v[GeV]$ & 40.6\\
\hline
$f_{33}v[GeV]$ & -48.2\\
\hline
$g_{12}v[GeV]$ & -2.06\\
\hline
$g_{13}v[GeV]$ & -1.77\\
\hline
$g_{23}v[GeV]$ & 0.427\\
\hline
$r_H$ & 78.0\\
\hline
$r_F$ & 146\\
\hline
$r_u$ & 0.190\\
\hline
$r_l$ & -9.19\\
\hline
$r_D$ & -6780\\
\hline
$r_R/v\times 10^{-10}$ & 2.27\\
\hline
\end{tabular}
\end{minipage}

\end{itemize}

\end{document}